\documentclass[structabstract]{aa}  
\usepackage{graphicx}
\usepackage{txfonts}
\begin{document}
   \title{A Renaissance study of Am stars}

   \subtitle{I. The mass ratio distribution}

   \author{Henri M.J. Boffin
                   }

   \institute{ESO, Alonso de C\'ordova 3107, Casilla 19001, Santiago, Chile\\
              \email{hboffin@eso.org}
                    }

   \date{Received ; accepted }

 
 \abstract
{} 
{Triggered by the study of Carquillat \& Prieur (2007, MNRAS, 380, 1064) of Am binaries, I reanalyse their sample of 60 orbits to derive the mass ratio distribution (MRD), assuming as they did a priori functional forms, i.e. a power law or a Gaussian. The sample is then extended using orbits published by several groups and a full analysis of the MRD is made, without any assumption on the functional form.}
{I derive the MRD using a Richardson-Lucy inversion method, assuming a fixed mass of the Am primary and randomly distributed orbital inclinations. Using the large sub-sample of double-lined spectroscopic binaries, I show that this methodology is indeed perfectly adequate.}
{I first derive new parameters of the functional form for the Carquillat \& Prieur sample. Using the inversion method, applied to my extended sample of 162 systems, I find that the final MRD can be approximated by a uniform distribution.}
   {}


   \maketitle

\section{Introduction}
Am stars are chemically peculiar A- or early F-type stars showing an overabundance of heavy elements and an underabundance of calcium and scandium. They were traditionally defined by the fact that the spectral types (ST) obtained using the metallic, the hydrogen and the calcium lines differ, such that ST(metal) $>$ ST(H) $>$ ST (Ca).
Such anomalies are convincingly explained as due to diffusion, more precisely by the competition between radiative and gravitational accelerations on atoms and ions in a relatively stable atmosphere (Michaud et al. \cite{Michaud83} and Talon, Richard, \& Michaud \cite{Talon06}). 

This gravitational settling can only be effective when the stars have a surface temperature between 7000 and 9000 K, and do not rotate too fast: their rotational velocities should be smaller than about 100--120 km/s, which is indeed observed for Am stars. Because contrarily to Ap stars, Am stars do not have external magnetic fields (Conti \cite{Conti}), the most likely mechanism to reduce high rotational velocities common to A-F stars is by tidal braking in binary systems (Roman, Morgan, and Eggen \cite{Roman48}). 

A number of observational studies have provided convincing evidence that the fraction of Am stars belonging to close binary systems with periods $P$ shorter than 100 days is relatively high  (e.g. Abt \& Levy  \cite{AL85}, Carquillat \& Prieur \cite{CP07}, Debernardi \cite{Deb02}). According to, e.g., Budaj (\cite{Budaj1997}), systems with $1 < P < 35$ days owe their slow rotation to tidal effects, either through synchronisation (for periods up to 12 days) or to pseudo-synchronisation (for eccentric systems). Systems with $P > 35$ days must be slow rotators from their formation on. Presumably these objects simply happen to be in the queue of the distribution of initial rotational velocities (e.g. Vuissoz \& Debernardi, \cite{VuiDeb}) or pre-main sequence tidal braking is the cause of their slow rotation (Abt \& Levy \cite{AL85}). An additional mechanism could be evolutionary expansion during the main sequence (Abt \& Levy \cite{AL85}).

Thus, Abt \& Levy (\cite{AL85}), following on previous studies made by Abt and colleagues, established  the frequency of binaries among Am stars. For an initial sample of 60 Am stars, they found 16 double-lined spectroscopic binaries (SB2), 20 single-lined spectroscopic binaries (SB1), and 20 visual and occultation companions not already counted as spectroscopic companions. This confirmed the high rate of binary systems among Am stars, giving weight to the binary explanation of the anomaly. Several other studies confirmed this. For example, Carquillat, Ginestet, \& Jaschek (\cite{Ca97}) studied a sample of 33 Am stars and detected a late-type companion for 22 systems from red spectra. Debernardi et al. (\cite{Debetal00}) studied Am stars in the Hyades and Praesepe open clusters. Of the 19 detected, they could ascertain the binary nature of 15. For the other four, they only have inconclusive evidence. The rate of Am binary is these clusters is thus at least of 79 \%, but the authors state that they cannot exclude that all are binaries.

In recent years, the study of Am stars and their binary properties has received a boost from several long-term studies. It is now possible to have a large sample of quality orbits of Am stars, allowing us to embark on a new, thorough study of the properties of these peculiar systems, which is what I call a Renaissance study. 

The trigger of the work was the study, performed over many years and entitled ``Contribution to the search for binaries among Am stars'', by J.-M. Carquillat and  J.-L. Prieur (Carquillat \& Prieur, \cite{CP07}; CP07 in the following). In this series of papers spanning almost a decade -- but based on data collected over more than 20 years -- the authors and various of their colleagues obtained information on a sample of 91 Am stars. They obtained elements for 60 orbits of 53 double or multiple spectroscopic binaries (SB), while in total 58 Am stars were identified as SB. Thus, their study confirms again that the rate of binary stars among Am stars is at least 64\%, larger than the already large rate of $47 \pm 3 \%$ found by Jaschek \& G\'omez (\cite{JaGo70}) for a sample of 295 normal A-type stars. In addition, 12 systems also belong to visual binaries, so that at least 70 out of their 91 Am stars are binaries. 
From their total sample of 60 orbits, they present a statistical study, including the mass ratio distribution (see next section).

Vuissoz \& Debernardi (\cite{VuiDeb}) presented a study of the mass ratio distribution of Am spectroscopic binary secondaries. They derived individual masses of 20 SB2 using the technique of Carrier et al. (\cite{Carrier}), based on photometric data, HIPPARCOS parallaxes, evolutionary tracks and observed mass ratios. They also derived the mass ratio distribution for an unpublished sample of over 200 spectroscopic binaries containing Am stars. For the SB1, they apparently used a Richardson-Lucy-like inverse algorithm using a constant mass for the primary star, verifying their method using their sample of 82 SB2. 
The mass ratio distribution for the whole (SB1+SB2) sample they derive is a gaussian-like curve centred around $q=0.5-0.$6, or, given the mass they assumed for the Am star, around secondary masses of 1.2--1.4 M$_\odot$. 
It is, however, unfortunate not to have more details on the sample used nor on the method of analysis. 

Another remarkable work, which is also at the core of the present paper, is the study done during the PhD of Debernardi (\cite{Deb02}; see also Debernardi, \cite{DebIAU}): an extensive survey of radial velocities of slowly rotating Am stars, using the spectrovelocimeter CORAVEL attached to the 1m Swiss telescope at the Observatoire de Haute-Provence, France. Of the 192 stars observed, at least 77\% of those which lie within 100 pc are binaries, with a high rate of SB2 systems, in stark contrast to CP07. Because both of these studies used CORAVEL, which was initially built to study cool stars, they are limited to  stars with rotational velocity $V\ \sin i \leq 45$ km/s. The radial-velocity precision -- and thus the rate of binaries -- depends also on the rotational velocity, being typically for Debernardi (\cite{Deb02}) larger than 0.5 km/s for slow rotators (10 km/s), and 4 km/s for rapid rotators (45 km/s). For CP07, the precision is about 0.5 km/s for slow rotators to about 1.5 km/s for the maximum rotational velocity limit of 40 km/s.

The present paper is organised as follow. In Sect. \ref{revcp07}, I revisit the analysis of CP07 and study the mass ratio distribution of their sample of 60 orbits. I will show that their estimates of the parameters of the functional forms they used are based on a wrong method. In order to derive the mass ratio distribution (MRD), I will use a proven inversion method. This will be applied to an extended sample of 162 orbits, as explained in 
Sect. \ref{extend}. Sect. \ref{mrd} will show the validity of the methodology used, and derive the MRD for the sub-samples of SB1 and SB2, while Sect. \ref{mrdall} will provide the final result as well as a discussion. 

\section{Revisiting the CP07 study}
\label{revcp07}
As mentioned above, one can only marvel at the long-term dedication of Carquillat, Ginestet, Prieur and their colleagues in obtaining the orbits of Am stars. In their 2007 paper, CP07 study the final catalogue of 60 orbits they obtain for their sample of 91 Am stars. In the present paper, I will concentrate on the study of the mass ratio distribution.

The 60 orbits of CP07 concern only 53 stars, as several are found to be multiple: they discovered five triple and two quadruple systems. They also have only 12 SB2, and among these, only one is a so-called twin, i.e. a system which has a mass ratio above 0.98. This very low number of SB2 may be due to the observing methodology, being for example against the findings of Abt \& Levy (\cite{AL85}; AL85 in the following) or Vuissoz \& Debernardi (\cite{VuiDeb}).

CP07 studied the mass distribution of the companion -- which is equivalent to the mass ratio distribution (MRD) as they fixed the primary mass -- based on the observed distibution of the mass function, $f(m)$, using a Monte Carlo method.
For a spectroscopic binary, the mass function is given by:
\begin{equation}
f(m) = \frac{q^3}{(1+q)^2} \ M_1 \ \sin ^3 i ,
\end{equation}
where $q=M_2/M_1$, with $M_1$ the mass of the Am primary star and $M_2$ the mass of the secondary, and $i$ is the (unknown) inclination angle of the orbit to the line-of-sight. As the Am stars are main-sequence stars and the primaries of these systems, the mass ratio should be smaller than or equal to one ($q \leq 1$).

CP07 tested two theoretical mass distributions: a power-law distribution $N(M_2) \propto M_2^{\alpha}$ and a Gaussian law, 
$$N(M_2) \propto exp {\left( - {\frac{(M_2-M_o)^2}{2\sigma _{\rm M} ^2}} \right)}.$$ 
The mass of the Am star was assumed to be fixed, $M_1=2$ M$_\odot$, and the mass of the secondary was truncated to 2 M$_\odot$ (since the companion is assumed to be less massive than the Am primary star). The assumption of a unique value of the mass of the primary can be considered appropriate given the limited range in effective temperature in which the Am phenomenon happens. Studies by Carquillat and Prieur of the individual systems provided them with $M_1$ in the range 1.6 to 2.2 M$_\odot$ (see e.g. table 11 in CP07). As the mass ratio depends, for low mass ratios, on the third root of $M_1$, this small range should not influence the general mass ratio distribution, if one uses a fixed primary mass.

\begin{figure}
   \centering
\includegraphics[width=8cm]{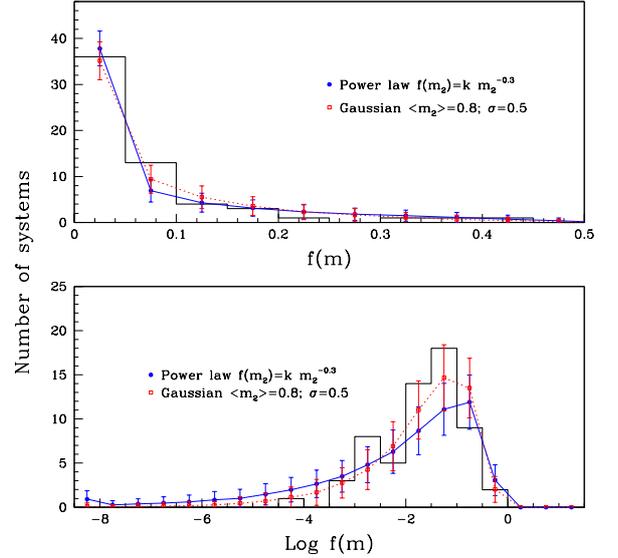}
      \caption{Comparisons between model and observed distributions for the sample of 60 SB from CP07. The upper panel shows the distribution of the mass function, $f(m)$, as well as the two best fits obtained by CP07: the solid line connecting the heavy dots with error bars corresponds to a power law, with index $ \alpha=-0.3$, while the dotted line connecting circles is a Gaussian, centred around 0.8 M$_\odot$ and with $\sigma _{\rm M} = 0.5~{\rm M}_\odot$. The corresponding distributions of $ \log f(m)$ are shown in the lower panel.
              }
         \label{Figfitcarq}
   \end{figure}

CP07 thus generated simulated samples whose distribution functions of $f(m)$ could be compared with the observed distribution of their sample of Am SBs. They then determined the best parameters for the power-law and Gaussian distributions by minimising the corresponding residuals, obtaining  $\alpha=-0.3 \pm 0.2$ for the power-law, and $M_o = 0.8$ M$_\odot$ with $\sigma _{\rm M}  = 0.5$ M$_\odot$ for the Gaussian distribution. These two distributions are in fact not so different from a uniform distribution: an $\alpha$ of 0 would indeed well be within the 2-$\sigma$ they obtained for the power law, while for the Gaussian distribution, the large $\sigma _{\rm M} $ compared to $M_o$ also makes for a very flat distribution. The fit to the $f(m)$ distribution, resulting from the two simulated distributions is shown in Fig.~\ref{Figfitcarq}. 

CP07 performed statistical tests to check the consistency of the two models with the observed sample of $f(m)$, which indicated that the fit is not significant for the two models and that both the power-law and the Gaussian distributions are indistinguishable. 

What is rather striking is that their distribution is very different from the one obtained by other authors. For example, Abt \& Levy (\cite{AL85}) derive the distribution of secondary masses $N = 6.39\ M_2^{0.84}$,
which is rather steep and explains the large number of SB2 they observe -- contrarily to CP07. Also, the Gaussian mass-ratio distribution  found by CP07 is centred around $q \simeq 0.4$, while Vuissoz \& Debernardi (\cite{VuiDeb}) derive a gaussian-like curve centred around q=0.5-0.6. It is of course possible that each of these surveys are different, but it may also hide something more fundamental. 

There is an important caveat, however, with the analysis of CP07 -- and of Vuissoz \& Debernardi (\cite{VuiDeb}) as well for that matter. They compared simulated and observed distributions of $f(m)$. This is very unfortunate, as is obvious from the upper panel of Fig.~\ref{Figfitcarq}: all the information is contained in only one (maximum two) bins, as these contains most of the systems when distributed linearly. It is thus rather difficult to make a good fit to the observed distribution. As shown by Boffin, Paulus  \&  Cerf (\cite{Bof92}) and Boffin, Cerf \& Paulus (\cite{Bof93}), in order to make a correct comparison, one should use the distribution\footnote{Of course, one can also use logarithmic spaced bins or, even better, have bins such that the number of systems in each bin is roughly the same.} of $\log f(m)$. This is shown in the lower panel of  Fig.~\ref{Figfitcarq}, where one can easily see that the fits are rather poor. The power law clearly makes a poor job, underestimating the number of systems with $-3 < \log f(m) < -1$, and overestimating the lowest part of the mass function. Although better, the Gaussian does neither provide a perfect match, showing that the parameters should be changed to improve the fit.

\begin{figure}
   \centering
   \includegraphics[width=8cm]{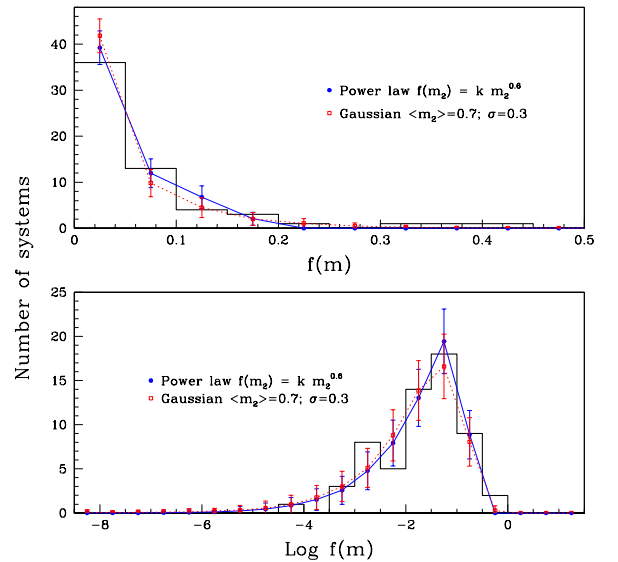}
      \caption{Same as Fig.~\protect{ \ref{Figfitcarq}}, with the new fit obtained in this paper, based on the distribution of $ \log f(m)$. In this case, the index of the power law is $ \alpha=0.6$, while the Gaussian is centred around 0.7 M$_\odot$ and has a $\sigma  _{\rm M} = 0.3~ {\rm M}_\odot$ (same symbols as before).  
              }
         \label{Fignewfitcarq}
   \end{figure}

I have thus redone their analysis, fitting the two distributions, a power-law and a Gaussian, so as to minimise the residuals between the simulated and observed distributions of $\log f(m)$. The results are shown in Fig.~\ref{Fignewfitcarq}. For the Gaussian case, I obtain a much more peaked distribution, slightly shifted, with $M_o=0.7, \sigma_M=0.3$, while for the power law, the results are completely different, as in order to have the best fit, I have to assume an {\bf increasing} function of the mass ratio, with $\alpha=0.6$, and that the mass of the secondary is smaller or equal to 1.25 M$_\odot$. These results, which also fit very well the distribution of $f(m)$ as one would expect, are in better agreement with the result of AL85. The fits shown in Fig.~\ref{Fignewfitcarq} are clearly very good, except -- for obvious reasons -- at the very high end of the mass function distribution. The truncated power law cannot be the 'real' distribution as the CP07 sample contains 12 SB2, 10 of which have $q > 0.626$, which are thus not represented if we cut $M_2$ at 1.25 M$_\odot$. I checked that to represent the observed distribution, one would indeed need to add just a few systems with $0.7 \leq q \leq 1.0$, to represent the SB2. In this and the previous figure, I have estimated the error bars of the simulations using 1000 Monte Carlo runs. Increasing this by a factor 10 had no effect.

One can also use the moments of the mass ratio distribution as computed from the distribution of $f(m)$ (Chandrasekhar \& 
M\"unch \cite{Cha50}), as explained by Heacox (\cite{Heacox95}). The moments of the 60 orbits of CP07 are then, using the coefficients of Heacox (\cite{Heacox95}): $<q>=0.426$ and $ <q^2>=0.23$. This would imply, naively, that a Gaussian fit would indeed be centred on $M_o=0.85$ M$_\odot$ and have $\sigma_M=0.42$ (assuming M$_1$=2 M$_\odot$). However, such a Gaussian would have a different mean, as negative masses (or mass ratios) are not allowed, nor are mass ratios larger than 1. This leads then to decrease $\sigma_M$ to 0.3 M$_\odot$ to have the correct mean. As explained above, the fit to the distribution of log $f(m)$ is better when also reducing $M_o$ to 0.7 M$_\odot$. This decreases the moments of the computed distributions, but they are still compatible -- within the error bars due to the limited sample -- with the observed ones, as confirmed by a Student test. The computed moments of the power law distribution for both values of $\alpha$ are also compatible with the observed moments. 

  \begin{figure}
   \centering
  \includegraphics[width=8cm]{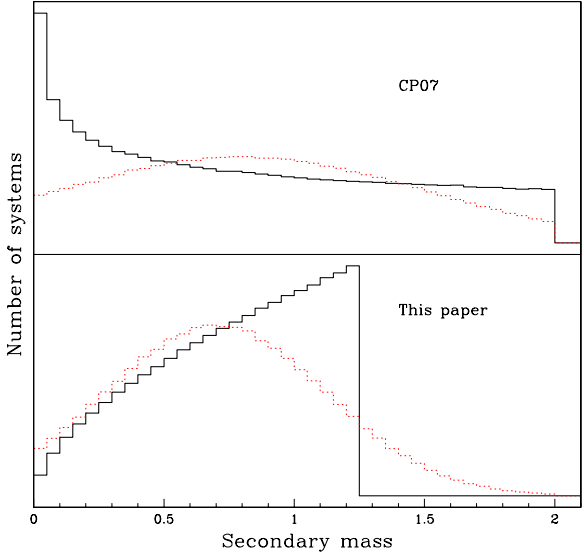}
      \caption{The distribution of the secondary masses for the sample of CP07. The upper panel shows the mass distribution obtained by these authors: either a power law (thick black line) or a Gaussian centred around 0.8 M$_\odot$ and with $\sigma  _{\rm M} = 0.5~{\rm M}_\odot$ (thin dotted red line). The two distributions are in fact very smooth and compatible with a uniform distribution. The bottom panel shows the distributions which best fit the sample of CP07 as obtained in this paper, fitting the distribution of $\log f(m)$ instead of $f(m)$. The resulting distributions are a truncated power law that increases with mass (thick black line) or a gaussian curve centred around 0.7 M$_\odot$ and with $\sigma  _{\rm M} = 0.3~ {\rm M}_\odot$ (thin dotted red line). 
              }
         \label{Figm2dis}
   \end{figure}
   
Similarly, one could check the validity of the distributions with a Kolmogorov-Smirnov test. CP07 already mentioned that their two distributions were not compatible with the observed $f(m)$ distribution, and this is indeed what I also found. The significance, however, greatly increases when using the distributions I derived, compared to the ones from CP07, as expected from seeing the better fits obtained. Finally, the distribution of the secondary mass obtained by CP07 and the one I obtained are shown in Fig.~\ref{Figm2dis}. 

Instead of assuming {\em a priori} a given functional form for the MRD, it would be better to {\em obtain} such an MRD from the observed distributions. This is what I will do in the next section, but in order to have a more statistically significant result, I will first extend the sample of Am stars with known orbits.

\section{Extending the sample}
\label{extend}
CP07 studied a sample of 60 orbits although the number of concerned Am stars is only 53. In fact, it may be misleading to use all the 60 orbits, assuming a constant mass of the primary corresponding to that of an Am star, when making a study of the mass ratio distribution. Indeed, for the triple or quadruple systems, one should use the {\bf total} mass of the binary system containing the Am star, and not the Am star alone. So, I used in the following only the 53 orbits corresponding to the ones most likely to contain the Am star (for quadruple systems) or the small period one (for triple systems) as this is most likely (but not neccesary) the reason why the rotation velocity of the Am star was reduced. Moreover, using data from the literature, it is possible to increase dramatically this sample.

CP07 already added to their sample of 60 orbits, 29 orbits from the 35 SBs of AL85  for which the quality of the orbit was good enough, and discarded the stars for which the orbital elements were qualified by the authors ``uncertain'' or ``marginal''. I have also added these 29 systems to my sample. Another large set of orbits can be found in the extensive work by Debernardi (\cite{Deb02}), who provides 70 orbits, as well as in Debernardi et al. (\cite{Debetal00}). Removing the duplicates among the orbits from the last two studies with these from CP07 and AL85, and further searching for orbits in the literature has allowed me to collect a final sample of 162 orbits. This is my new, extended sample, whose diverse origin allows me to think that it is not biased and a statistical analysis of it would be relevant. Most important is that this sample almost doubles the extended sample of CP07 and is therefore much more representative. In my search in the literature, I did not include systems for which the Am star is only the secondary, as this would invalidate the current procedure. 

A full presentation of my extended sample is deferred to another paper. I just note here that it contains 98 SB1 and 64 SB2. The fraction of SB2 (40 \%) is thus close to the one found by AL85 (44 \%), but much larger than the one found by CP07 (20 \%).
In the next section, I analyse the MRD distribution as obtained from this new sample.  I do not expect necessarily to obtain the same MRD as the one derived from the CP07 sample, as the larger presence of SB2 implies a larger fraction of high mass ratio systems.

\section{The mass ratio distribution of SB1 and SB2}
\label{mrd}
\subsection{Double-lined spectroscopic binaries as testbed}

\begin{figure}
   \centering
  \includegraphics[width=8cm]{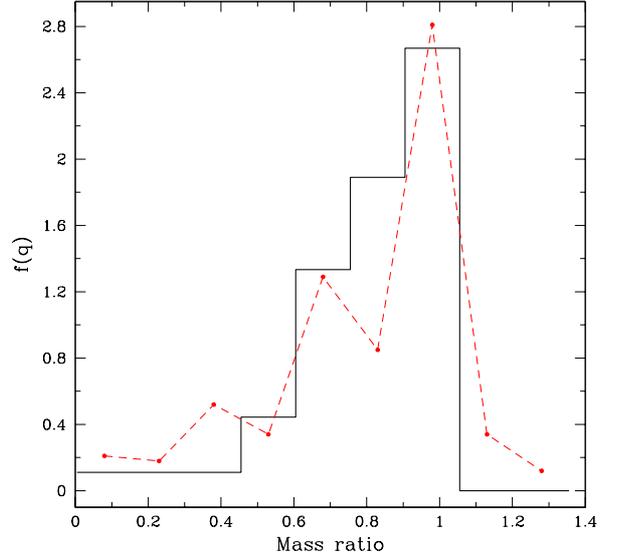}
      \caption{The observed mass ratio distribution for a sub-sample of 60 SB2 systems studied in this paper (black histogram) is compared to the distribution derived using a Richardson-Lucy inversion method (solid dots connected with the dashed red line). 
              }
         \label{Figsb2fq}
   \end{figure}

  \begin{figure}
   \centering
  \includegraphics[width=8cm]{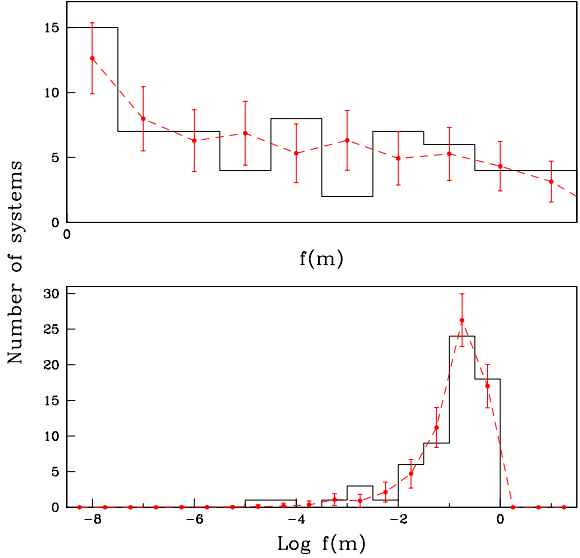}
      \caption{Comparison between the observed distribution of the  mass function (upper panel) of the sub-sample of SB2 and of its logarithm (lower) with the computed distributions, using the observed mass ratios (see Fig.~\protect{ \ref{Figsb2fq}}), a primary mass of 2 M$_\odot$, and a random inclination. 
              }
         \label{Figsb2fm2}
   \end{figure}

The sub-sample of 64 double-lined spectroscopic binaries is of course of crucial importance, as it will allow me to thoroughly test the method I plan to use to derive the MRD. This is done following Boffin et al. (\cite{Bof93}) and Cerf \& Boffin (\cite{CB94}), \emph{i.e.} using a Richardson-Lucy inversion method. I refer to the two above-mentioned papers for further details of the method and just note that I always stopped after 5 iterations, to avoid the solution becoming too fine-grained. I also used bins of 0.15 in the mass ratio, so as to have a statistically significant solution. 

The SB2 provide of course a direct estimate of the mass ratio, and its distribution is shown as an histogram in Fig.~\ref{Figsb2fq}. Note that 4 of the 64 SB imply a mass ratio above 1, which are hard to explain in systems where the Am star is the primary. The above unity value is most likely due to observational errors and, instead of arbitrarily assuming a mass ratio of one, they were not considered in the following analysis. The observed MRD for the SB2 is an increasing function of the mass ratio. It is noteworthy that an SB2 can still be detected with a mass ratio as small as $~0.5$, but of course, an SB2 will be more easily discovered when $q$ is close to one. It is not obvious to disentangle the observational bias from the original distribution, and this is why one should study a sample of SB1 and SB2, as I will do below. However, I can now use the SB2 to test the assumptions and method used in this paper. Fig.~\ref{Figsb2fq} also shows the MRD I obtained using the Richardson-Lucy inversion method. The good match between the calculated and real distributions shows the validity of the inversion procedure which can thus be applied to the SB1 systems in the sample.

The next things to verify is whether the methodology used in the previous section is valid. For this, I used the MRD as observed for the sub-sample of SB2, and then ran Monte Carlo simulations, using a constant mass for the primary, $M_1=2$ M$_\odot$, and a random inclination to compute a simulated distribution of  $\log f(m)$,\emph{ i.e.} I repeated the procedure of Sect. \ref{revcp07}. If the assumptions hold, then the deduced simulated distribution should be compatible with the observed one. The results, which are positive and convincing, are shown in 
Fig.~\ref{Figsb2fm2}. Thanks to this control sample of SB2, we can now be confident that both the Richardson-Lucy inversion and the assumptions made are valid, and I can thus apply this to the sample of single-lined spectroscopic binaries.

  \begin{figure}
   \centering
  \includegraphics[width=8cm]{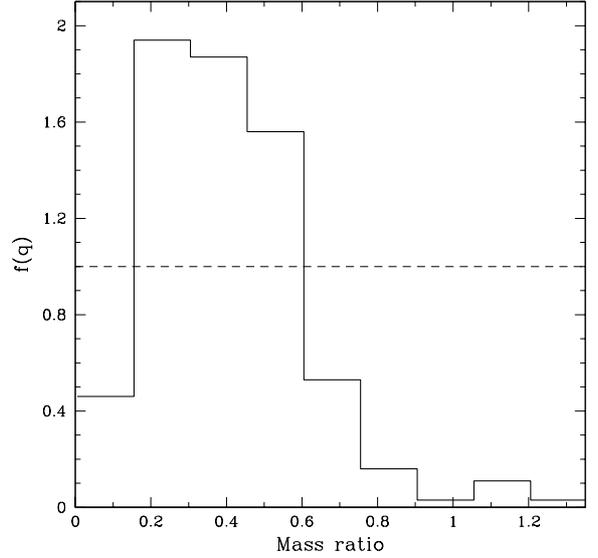}
      \caption{The mass ratio distribution for the sub-sample of 98 SB1 systems studied in this paper, derived using a Richardson-Lucy (R-L) inversion method. The distribution is compatible with a uniform distribution for $q < 0.6$. 
      The dashed line corresponding to a uniform distribution over the whole range of mass ratios is shown for comparison.
              }
         \label{Figsb1fq}
   \end{figure}

  \begin{figure}
   \centering
  \includegraphics[width=8cm]{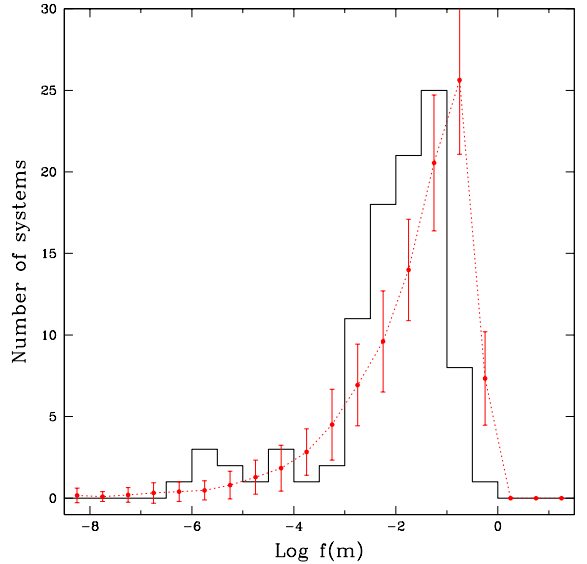}
      \caption{The distribution of the logarithm of the mass function for the sub-sample of SB1 (black histogram) is compared to the one obtained by assuming that the mass ratio distribution is constant (heavy dots with error bars connected by the thick line). The errors bars are the results of 1000 Monte Carlo simulations.
              }
         \label{Figsb1fmfqcst}
   \end{figure}

 \subsection{Single-lined binaries}
My extended sample contains 98 SB1, for which I can apply the Richardson-Lucy inversion method, assuming a constant mass of the primary and a random inclination angle. The result of this inversion is shown in  Fig.~\ref{Figsb1fq}. The sample of SB1 does not contain many systems with $q > 0.65$, which agrees with the fact that such systems would appear as double-lined, as deduced from the SB2 MRD. Within the range $0.15 < q < 0.6$, the obtained MRD is not very different from a uniform distribution. The deficiency of systems with $q < 0.15$ is due to a combination of observational bias and a bias of the inversion method (see, e.g., Mazeh \& Goldberg \cite{Mazeh}, Heacox \cite{Heacox95}). 

To make sure that one cannot assume a constant MRD over the whole range of mass ratios, $0 < q < 1$, I have calculated the distribution one would obtain in this case, and compared this with the observed distribution of  $\log f(m)$ for the sub-sample of SB1. The result, shown in Fig.~\ref{Figsb1fmfqcst}, confirms that a constant MRD over the whole range of mass ratios is not compatible with the observation, in particular, due to the overestimate of systems with $\log f(m) > -1$. This is in line with the fact that a system with too large a mass ratio would appear as a SB2, not as an SB1. It remains thus to be seen what happens when looking at the 
whole sample of SB1 and SB2. This is what I set about doing in the next section.

\section{The MRD of Am stars}
\label{mrdall}
\begin{figure}
   \centering
 \includegraphics[width=8cm]{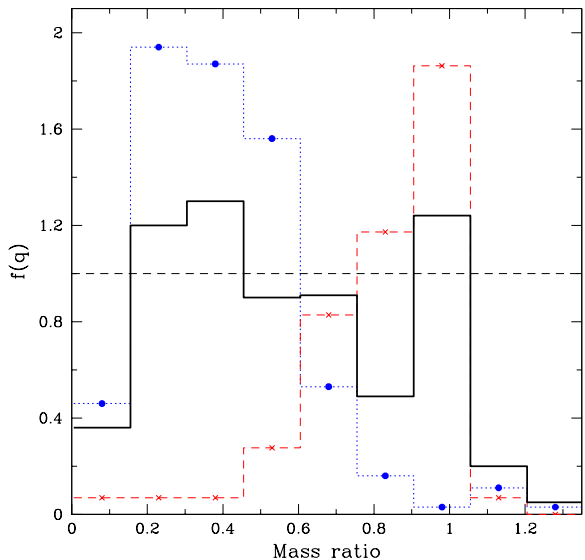}
      \caption{The mass ratio distribution for the whole sample of 162 systems studied in this paper. The dashed red histogram corresponds to the observed distribution for SB2 systems, while the blue dotted one is the $q$-distribution obtained from a Richardson-Lucy (R-L) inversion of the sub-sample of SB1 systems. The Am star mass ratio distribution is thus the sum of the two distributions. The black heavy line shows the mass ratio distribution as derived from a R-L inversion of the whole sample of SBs, giving a distribution which is comparable to half the sum of the two distributions.
              }
         \label{Figfqsb1andsb2}
   \end{figure}

To determine the mass ratio distribution of the whole sample of Am stars, comprising 162 systems among which 98 SB1 and 64 SB2, there are two ways. The first one is simply to combine the results of the two previous subsections, that is, combine the {\em observed} MRD of SB2 with the {\em deduced} MRD of SB1. This is shown in Fig.~\ref{Figfqsb1andsb2}, where one can see that these two MRDs complement each other. Another way to compute the MRD of the whole sample is to work with the SB2 the same way as with the SB1, \emph{i.e.}  create a full sample of $f(m)$ and invert it using our Richardson-Lucy algorithm. This is also shown in Fig.~\ref{Figfqsb1andsb2} and confirms that both methods provide similar answers.

  \begin{figure}
   \centering
  \includegraphics[width=8cm]{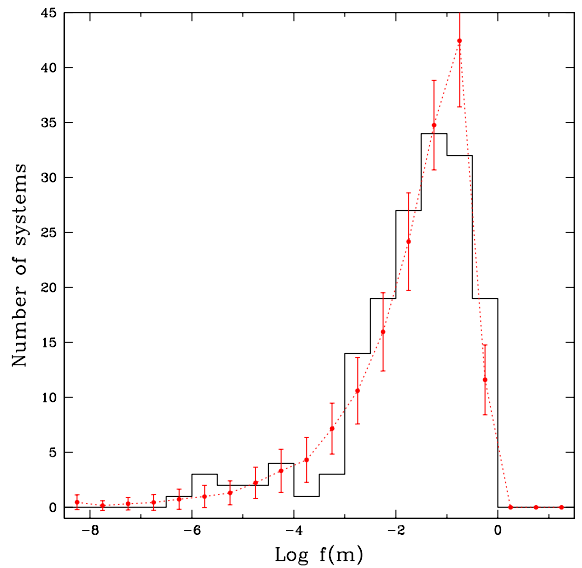}
      \caption{The distribution of the logarithm of the mass function for the whole sample of spectroscopic binaries (black histogram) is compared to the one obtained by assuming that the mass ratio distribution is constant (red solid line with error bars). The errors bars are the results of 1000 Monte Carlo simulations. 
              }
         \label{Figfmwithfqcst}
   \end{figure}

Apart from the understood deficit of systems at very low $q$, the final MRD that I obtain seems to show slight hints for a double-peaked distribution, with a broad peak around $q \sim 0.3$ and one, narrow, close to $q \sim 1$. The moments as derived from the $f(m)$ values are: $<q>=0.52, \, <q^2>=0.37$.

However, the difference with a constant MRD is not huge and such a simple MRD could well be considered. To prove this point, I compare in 
Fig.~\ref{Figfmwithfqcst} the observed distribution of $\log f(m)$ with that obtained when assuming that $M_2 = 2 $\ M$_\odot$, $f(q) = 1$, and $f(i) \propto \sin i$. Apart from a possible small overestimate of the computed distribution in the bin with $-1 < \log f(m) < -0.5$ (possibly indicating that the deficiency of systems around $q \sim 0.8$ is significant), the two distributions agree pretty well with each other, and one should perhaps not try to look much further for the real distribution.

It may be useful to see how this distribution compares with ones obtained for other samples. I have  shown earlier (\cite{SB9}) that the mass ratio distribution of the binary systems with main sequence primaries in the SB9 catalogue\footnote{http://sb9.astro.ulb.ac.be/} depends on the spectral type of the primary: there is an apparent continuous trend from massive primaries (having a MRD peaked to lower mass ratios) to low-mass primaries (K stars being compatible with an uniform MRD). The SB9 catalogue is however possibly subject to severe observational biases and so this result would need a careful confirmation. \cite{Goldberg} made a study of a complete sample of 129 SB with periods below 2500 days, for which the primary mass is known and is between 0.5 and 1 M$_\odot$ (i.e. mostly G-K dwarfs). They find a distribution that seemingly includes two ''populations'', one with a high asymmetric peak at $q \sim 0.2$ and another with a smaller peak at $q \sim 0.8$, while the minimum between the two populations is centered at $q \sim 0.55$. However, the distribution seems to be different between halo and disc stars, as well as between ''high'' ($M_1 > 0.67$ M$_\odot$) and ''low'' mass primaries.
\cite{Halbwachs} derived the statistical properties of 89 main-sequence binaries with spectral types F7 to K and with periods up to 10 years. Their distribution of mass ratios presents two maxima: a broad (flat) peak from $q \simeq 0.7$, and a sharp peak for $q > 0.8$. Their MRD also depends on the orbital period, since short period binaries ($P < 50$ days) include more systems with mass ratios of 0.8 or more.
\cite{Mazeh2003} reported infrared spectroscopic observations of a large well-defined sample of main-sequence, single-lined spectroscopic binaries to detect the secondaries and derive the mass ratio distribution of short-period binaries. Their sample consists of 51 Galactic disc spectroscopic binaries, with primary masses in the range 0.6--0.85 M$_\odot$. Their mass ratio distribution is approximately constant over the range of $q$ between 0.3 and 1. The distribution appears to rise at lower $q$ values, but the uncertainties are
sufficiently large that they cannot rule out a distribution that remains constant over the full range of mass ratios.
\cite{Fisher} obtained the mass ratio distributions of the local population of field
binaries by studying a volume-limited sample of 371 spectroscopic binaries in the solar neighbourhood. Using a Monte-Carlo method, they obtain an MRD which is almost uniform for $q < 0.8$ and a peak towards $q = 1$ (SB2 systems).
Very recently, \cite{Raghavan} analysed the properties of companions to solar-
type stars in a sample of 454 stars. They derived a roughly flat distribution
for mass ratios between 0.2 and 0.95, with an excess of twins, which prefer relatively short orbital periods. The MRD I derive in the present paper is thus compatible with most of these studies. 

Perhaps even more interesting is to note that the MRD obtained for G-K giants member of spectroscopic binaries (Boffin et al. \cite{Bof92},  \cite{Bof93}) is also compatible with a flat distribution. As Am stars will evolve into G-K giants, it is thus reassuring that these two populations seem to share the same MRD.

Because the primaries have all about the same mass, 2 M$_\odot$, the MRD represents also the mass distribution of the secondaries. The uniform distribution that I derive is very different from the ubiquituous Initial Mass Function observed for the distribution of the mass of single stars, and which give a large preference to low-mass stars (\cite{Salpeter}, \cite{Kroupa}). The Am systems -- and the many other samples described above -- have thus most likely not been assembled by random pairing of stars from an IMF.

Fig.~\ref{Figfqsb1andsb2} shows that the real issue in obtaining the final MRD of the Am binaries is to establish the correct relative fraction of SB1 and SB2. Observational biases could influence this ratio, such as the \"Opik or Branch effect (AL85), and it is not an easy task to disentangle those (Mazeh \& Goldberg \cite{Mazeh}). However, as the current sample is a collection of several samples, which seem to complement each other very well and which must all have been affected in different -- hopefully opposite -- ways, and as they were not magnitude limited, I think it is reasonable that the sample is well representative and the conclusions on the MRD should be secure.The final word on this will only be said once we have a still much large, controlled sample of Am stars with known orbital elements. This is not for the immediate future I am afraid.

\section{Conclusions}
In this Renaissance study of Am stars, I have reanalysed the distribution of the mass ratio distribution -- or equivalently, as the mass of the Am stars can be assumed constant, the distribution of the secondary masses -- using a rigourous approach and an extended sample.
 \begin{enumerate}
      \item Reanalysing the sample of 60 orbits from CP07, I have shown that, unfortunately, their conclusion on the MRD is not correct, as they did not fit the distribution of $\log f(m)$. A proper analysis shows that a power law with a {\em positive} index of 0.6 and assuming that $q \lesssim 0.6$ is a good fit to their sample, which is characterised by a small fraction of double-lined spectroscopic binaries. If one would like to seek a Gaussian function, the best fit is given with $M_o = 0.7$ M$_\odot$, and $\sigma_M = 0.3$ M$_\odot$.  
        \item I have extended the sample of Am binaries with precise orbital elements, collecting the results of several surveys, as well as other results from the literature. This new, extended sample contains 162 systems, of which 98 SB1 and 64 SB2. 
      \item Using the SB2 sub-sample, I have shown that it is correct to use a Richardson-Lucy inversion method, assuming in order to derive the MRD a constant mass for the Am primary star as well as random inclinations.
      \item Based on this, I have estimated the MRD for the whole sample. The final MRD obtained seems to slightly hint at a double-peaked distribution, with a broad peak around $q \sim 0.3$ and one close to $q \sim1$, although a flat MRD seems to be a very good fit as well. As always, the final shape of the MRD depends, however, on the exact ratio between single and double-lined spectroscopic binaries.
  \end{enumerate}
   
\begin{acknowledgements}
      It is a great pleasure to thank former ESO librarian Chris Erdmann -- now at the Harvard-Smithsonian Center for Astrophysics -- for his very efficient help in getting some of the references used in this work. My work has made use of the SIMBAD database,
operated at CDS, Strasbourg, France, as well as NASA's Astrophysics Data System. 
      I also wish to thank my kids and wife for their patience while I was working on this paper, instead of sharing valuable moments with them. I thank the referee, Helmut Abt, and the editors for their swift and efficacious handling of the manuscript.
\end{acknowledgements}

\end{document}